\documentclass[12pt]{book}

\usepackage[dvips]{graphicx,color}
\usepackage{makeidx,tsukuba}

\makeauthorindex
\makeindex

\begin{document}

\BookTitle{\itshape The 28th International Cosmic Ray Conference}
\CopyRight{\copyright 2003 by Universal Academy Press, Inc.}
\pagenumbering{arabic}

\chapter{
 A New Estimate of the Extragalactic Gamma-Ray Background from EGRET Data}

\author{%
Andrew W.~Strong,$^1$ Igor V.~Moskalenko,$^{2,3}$ and Olaf Reimer$^4$ \\
{\it (1) Max-Planck Institut f\"ur extraterrestrische Physik, Garching, Germany\\
(2) NASA/Goddard Space Flight Center, Code 661, Greenbelt, MD 20771, USA\\
(3) JCA/University of Maryland, Baltimore County, Baltimore, MD 21250, USA\\
(4) Institut f\"ur Theoretische Physik IV, Ruhr-Universit\"at Bochum, Germany
} \\
}

\section*{Abstract}
We use the GALPROP model for cosmic-ray (CR) propagation to obtain our best
prediction of the Galactic component of gamma rays, and show that away
from the Galactic plane it gives an accurate prediction of the observed
EGRET intensities. On this basis we re-evaluate the isotropic
extragalactic gamma-ray background (EGRB). We find that for some energies
previous work underestimated the Galactic contribution and hence
overestimated the background. Our new EGRB spectrum shows a
positive curvature similar to that expected for models of the
extragalactic emission based on the blazar population.

\section{Introduction}

The GALPROP model for CR propagation produces explicit
predictions for the angular distribution of diffuse Galactic
$\gamma$-rays.
In a companion paper [4] we describe a comparison of a model for 
Galactic diffuse continuum $\gamma$-rays with EGRET data. 
In this paper we use this model to determine the EGRB.

\section{Data and method}
Given the success of the model [4] in reproducing the $\gamma$-ray sky, we can
use it to make improved estimates of the EGRB. Since the model is
nevertheless not exact the best approach is to fit the observed
intensities with a free scaling factor; in this way the EGRB is determined
as the intercept, thus removing any residual uncertainty in the
absolute level of the Galactic components. (This is the same  method
as in [2], the difference lies in the model).  To reduce
the effects of Galactic structure, point sources etc.\ the fits are made
excluding the plane, using the region 
$360^\circ<l<0^\circ,10^\circ<|b|<90^\circ$ (``region G''); ideally both IC and
gas-related components would be left free but they are difficult to
separate statistically at high latitudes, so we make a linear fit to
the total IC+$\pi^0$-decay+bremsstrahlung, with the scaling factor and EGRB
as parameters. The fit and errors are based on a simple $\chi^2$ analysis,
with ($l,b$) bins $360^\circ\times2^\circ$ to obtain sufficient statistics
(at least 10 counts per bin were required).
For comparison we also made fits to the entire sky; in
this case IC and gas-related contributions are easily separated; the
two fit regions then give some indication of the model-dependent
systematic error in our EGRB estimates.

The scaling factors determined for the region G fits 
reflect the deviations from the model and are shown in Table 1.
They are typically between 0.8 and
1.2 which is satisfactory. (Only statistical errors are shown in Table 1.)
For 30--50 MeV, 1--2 and 2--4  GeV the
scaling factors deviate further from 1 reflecting the discrepancy in
the spectrum
so that these are the least reliable ranges of our EGRB determination.
The EGRB is however not very sensitive to the scaling factor.


Table 2 summarizes the fitted EGRB values.
The two fitted regions (region G and
all sky) give consistent results, indicating that there is no large
systematic effect; it shows a model-dependent systematic uncertainty of 5--30\%,
comparable to  the formal statistical errors.  This is comparable to
the $\sim15$\% systematic uncertainty on EGRET data so we adopt 30\%
for our total error estimate. 

Fig.\ 1 shows the extragalactic X- and $\gamma$-ray background, 
using the compilation by [2] but with our new EGRET values, and
also updated COMPTEL results ([5] and references therein).
Our estimates lie significantly below those of [2], in
each energy range.  The positive curvature in our EGRB spectrum is
interesting and not unexpected [1] but in view of the
systematic uncertainties should not be taken too literally; a similar,
less pronounced effect is present in the Sreekmar spectrum.
A power-law is a poor fit to our spectrum. 
 Although the  50 MeV -- 2 GeV range can be fit
satisfactorily by a power law, it is clearly inconsistent with the points above 2 GeV.

The reason for the difference between our spectrum and that given in [2]
is the improved modelling of high-latitude $\gamma$-rays
based on inverse Compton emission from the halo. An indication of this effect is already apparent in Table 1 of [2]
which requires scaling factors up to 1.8 above 1 GeV. 
Detailed comparisons of our model with the EGRET data will be given
in a forthcoming publication.


I.V.M.\ acknowledges partial support from
a NASA Astrophysics Theory Program grant.

\begin{table}[!t]
\caption{Scaling factors of components of Galactic diffuse emission}
\label{scaling_factors_EB}
\begin{center}
\begin{tabular}{rrrr}
\hline
 &&\multicolumn{2}{c}{All-sky}\\
\cline{3-4}
Energy, MeV & Total, $|b|>10^\circ$&IC\hspace*{5ex} &Gas\hspace*{3ex}\\
\hline
$  30-50$~~~~&$1.43\pm0.05$~~~~&$1.54\pm0.12$~~&$1.20\pm0.14$   \\
$  50-70$~~~~&$1.15\pm0.03$~~~~&$1.10\pm0.065$~&$1.03\pm0.05$   \\
$  70-100$~~~&$1.07\pm0.02$~~~~&$1.20\pm0.04$~~&$0.82\pm0.02$   \\
$ 100-150$~~~&$0.97\pm0.01$~~~~&$1.14\pm0.03$~~&$0.76\pm0.01$   \\
$ 150-300$~~~&$0.93\pm0.01$~~~~&$1.16\pm0.03$~~&$0.76\pm0.01$   \\
$ 300-500$~~~&$1.07\pm0.01$~~~~&$1.26\pm0.04$~~&$0.90\pm0.01$   \\
$ 500-1000$~~&$1.17\pm0.015$~~ &$1.31\pm0.04$~~&$1.06\pm0.01$   \\
$1000-2000$~~&$1.34\pm0.025$~~ &$1.21\pm0.06$~~&$1.37\pm0.03$   \\
$2000-4000$~~&$1.48\pm0.05$~~~~&$1.14\pm0.08$~~&$1.67\pm0.06$   \\
$4000-10000$~&$0.83\pm0.06$~~~~&$0.58\pm0.09$~~&$1.40\pm0.11$   \\
\hline
\end{tabular}
\end{center}
\end{table}

\begin{table}[!t]
\caption{Estimates of EGRB$^1$ obtained by fitting optimized model to EGRET data}
\label{estimates_of_EB}
\begin{center}
\begin{tabular}{rrrr}
\hline
Energy, MeV\hspace*{1ex}&Total,$^2$ $|b|>10^\circ$&All-sky,$^2$ IC+Gas& [2]$^3$\hspace*{6ex}\\
\hline
$  30-50$~~~~&$16.4~\pm0.67$~ &$16.2~~\pm0.8$~~~&$24.0~~~\pm7.0$~~~~\\
$  50-70$~~~~&$10.2~\pm0.19$~ &$10.6~~\pm0.2$~~~&$13.3~~~\pm2.6$~~~~\\
$  70-100$~~~&$6.33~\pm0.10$~ &$ 6.42~\pm0.1$~~~&$ 7.83~\pm1.05$~~ \\
$ 100-150$~~~&$4.26~\pm0.07$~ &$ 4.3~~\pm0.07$~ &$ 5.5~~\pm0.75$~~ \\
$ 150-300$~~~&$3.76~\pm0.06$~ &$ 3.76~\pm0.07$~ &$ 5.4~~\pm0.72$~~ \\
$ 300-500$~~~&$1.08~\pm0.04$~ &$ 1.1~~\pm0.04$~ &$ 1.97~\pm0.27$~~ \\
$ 500-1000$~~&$0.65~\pm0.04$~ &$ 0.67~\pm0.04$~ &$ 1.36~\pm0.19$~~ \\
$1000-2000$~~&$0.265\pm0.03$~ &$ 0.33~\pm0.03$~ &$ 0.617\pm0.084$ \\
$2000-4000$~~&$0.203\pm0.02$~ &$ 0.28~\pm0.02$~ &$ 0.30~\pm0.044$ \\
$4000-10000$~&$0.117\pm0.02$~ &$ 0.125\pm0.02$~ &$ 0.196\pm0.029$ \\
\hline
\end{tabular}
\end{center}
\small
$^1$Units: $10^{-6}$ cm$^{-2}$ sr$^{-1}$ s$^{-1}$. \\
$^2$For our fits only statistical 
errors are given, for systematic errors see text.\\
$^3$The values [2] are from their Table 1 which includes systematic errors.
\end{table}

\section{References} 
\re
1.\ Salamon M.H., Stecker F.W.\ 1998, ApJ 493, 547
\re
2.\ Sreekumar P.\ et. al.\ 1998, ApJ 494, 523
\re
3.\ Strong A.W., Moskalenko I.V., Reimer O.\ 2000, ApJ 537, 763 
\re
4.\ Strong A.W., Moskalenko I.V., Reimer O.\ 2003, these Proceedings
\re
5.\ Weidenspointner G.\ et al.\ 2000, in AIP Conference Proceedings 510, 467

   \begin{figure}[!tb]
   \centering
    \includegraphics[height=1\textwidth,angle=90]{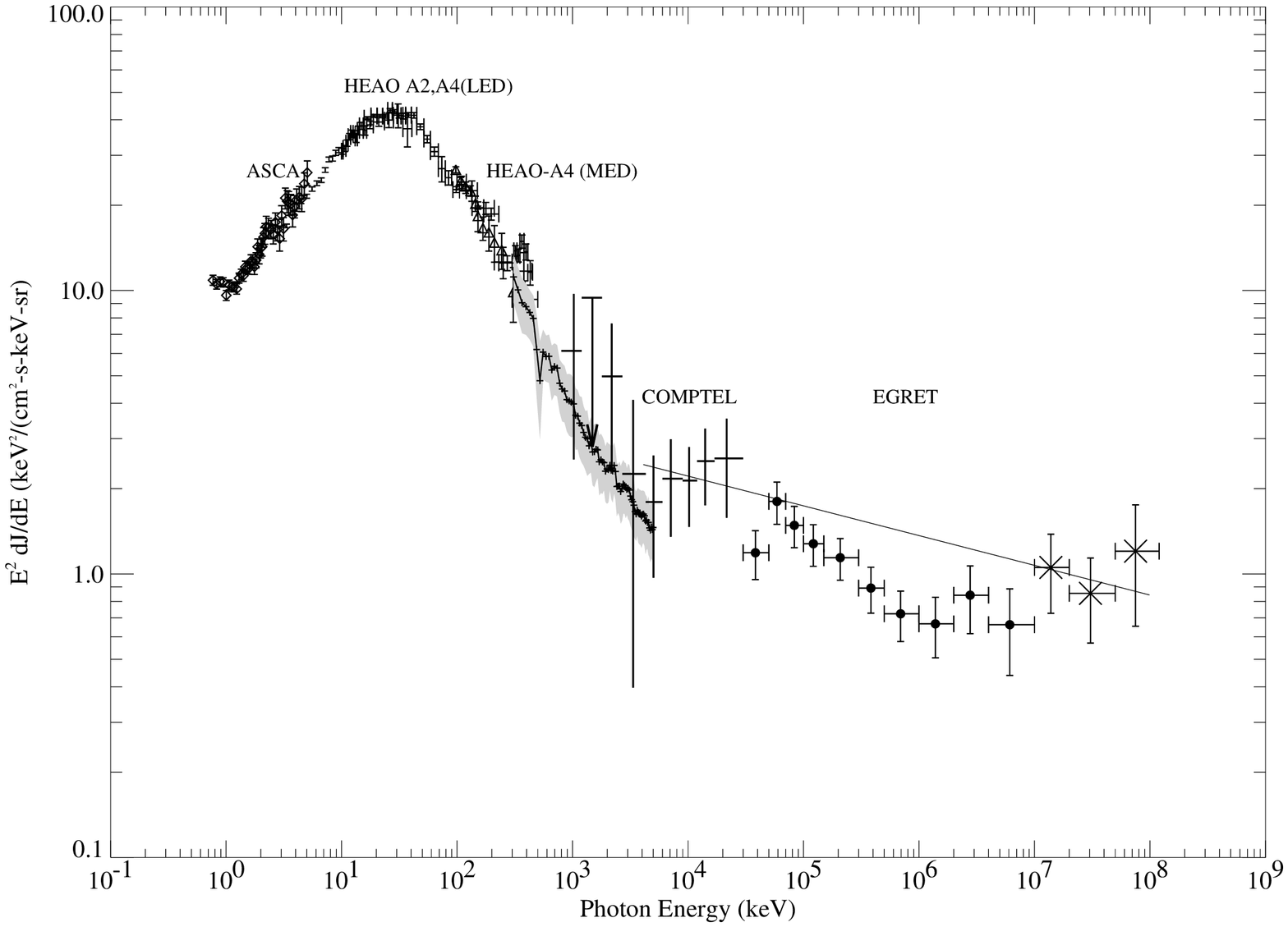}
     \caption{
Extragalactic X-ray and $\gamma$-ray spectrum. 
Data compilation from  [2]  except for COMPTEL [5] and EGRET 30 MeV -- 10 GeV (this work).
The power-law in the EGRET range and the 3 data points above 10 GeV represent 
the spectrum derived by [2].  }        
         \label{spectrum_EXGRB}
   \end{figure}

  \endofpaper
\end{document}